\documentclass[aps,pra,showpacs,groupedaddress,floatfix,nofootinbib]{revtex4}
\usepackage{amsmath}

\usepackage{graphicx}
\usepackage{dcolumn}

\begin{document}

\title{On two exactly-solvable one-dimensional Hamiltonians with PT symmetry}

\author{Francisco M. Fern\'andez}\email{fernande@quimica.unlp.edu.ar}

\affiliation{INIFTA (UNLP, CCT La Plata-CONICET), Blvd. 113 y 64 S/N, \\
Sucursal 4, Casilla de Correo 16, 1900 La Plata, Argentina}

\begin{abstract}
We provide an explanation to the behaviour of the spectra of two
exactly-solvable one-dimensional Hamiltonians with PT symmetry proposed
earlier. We calculate the branch points at which pairs of eigenvalues
coalesce and discuss the perturbation series.
\end{abstract}


\maketitle

\section{Introduction}

\label{sec:intro}

Some time ago Ahmed\cite{A09} solved the Schr\"{o}dinger equation with the
PT-symmetric potentials $V_{1}(x)=x^{2}/4+ig|x|x$ and $V_{2}(x)=|x|+igx$
exactly and found an interesting behaviour of the corresponding point
spectra. In both cases there is an infinite number of real eigenvalues when $%
g=0$ and as $g$ increases the number of real eigenvalues decreases to just
one. At every finite nonzero value of $g$ there is an odd number of real
eigenvalues (say $2k+1$) and as $g$ increases this number reduces to $%
2k-1,\,2k-3,2k-5,\ldots ,1$. Ahmed did not explain this phenomenom
satisfactorily and did not identify the critical $g$ values at which each
change takes place. However, he proposed to call it \textit{scarcity of real
discrete eigenvalues}.

The purpose of this paper is to provide a more detailed discussion of the
behaviour of the spectra of those models. In section~\ref{sec:perturbation}
we outline some properties of the perturbation series for a particular class
of parameter-dependent Hamiltonians and in section~\ref{sec:models} we
discuss the spectra of the models mentioned above. Finaly, in section~\ref
{sec:conclusions} we draw conclusions.

\section{Perturbation series}

\label{sec:perturbation}

Let $H(\lambda )$ be a parameter-dependent linear operator and $U$ a linear
invertible operator such that $UH(\lambda )U^{-1}=H(-\lambda )$. Therefore,
if $\psi _{n}(\lambda )$ is an eigenfunction of $H(\lambda )$ with
eigenvalue $E_{n}(\lambda )$ then it follows from
\begin{equation}
UH(\lambda )\psi _{n}(\lambda )=UH(\lambda )U^{-1}U\psi _{n}(\lambda
)=H(-\lambda )U\psi _{n}(\lambda )=E_{n}(\lambda )U\psi _{n}(\lambda ),
\label{eq:UHpsi}
\end{equation}
that $U\psi _{n}(\lambda )$ is proportional to an eigenfunction $\psi
_{m}(-\lambda )$ of $H(-\lambda )$ with eigenvalue
\begin{equation}
E_{m}(-\lambda )=E_{n}(\lambda ).  \label{eq:Em(-l)=En(l)}
\end{equation}
Here, we are interested in the case that $H(0)$ is an Hermitian operator.
Since equation (\ref{eq:Em(-l)=En(l)}) is assumed to be valid for all $%
\lambda $ we conclude that $E_{m}(0)=E_{n}(0)$. Consequently, if the
eigenvalues of $H(0)$ are nondegenerate then $m=n$ and $E_{n}(\lambda
)=E_{n}(-\lambda )$ for all $\lambda $.

If the eigenvalue $E_{n}(\lambda )$ can be expanded in a Taylor series about
$\lambda =0$
\begin{equation}
E_{n}(\lambda )=\sum_{j=0}E_{n}^{(j)}\lambda ^{j},
\label{eq:En_lambda_series}
\end{equation}
then $E_{n}^{(2j+1)}=0$ for all$\;j=0,1,\ldots $. If this expansion
converges for $|\lambda |<R_{n}$ then
\begin{equation}
E_{n}(ig)=\sum_{j=0}E_{n}^{(2j)}(-g^{2})^{j},  \label{eq:En_g_series}
\end{equation}
is real for all $|g|<R_{n}$ provided that $g$ and the perturbation
coefficients $E_{n}^{(2j)}$ are real.

\section{Two exactly-solvable PT-symmetric models}

\label{sec:models}

As indicated in the Introduction Ahmed\cite{A09} briefly discussed two
exactly-solvable one-dimensional PT-symmetric models. One of them is given
by
\begin{equation}
H(\lambda )=p^{2}+\frac{1}{4}x^{2}+\lambda |x|x,  \label{eq:H_model1}
\end{equation}
where, $[x,p]=i$. In this case the parity transformation
\begin{equation}
PpP=-p,\;PxP=-x,  \label{eq:parity_transf}
\end{equation}
yields $PH(\lambda )P=H(-\lambda )$, where $P^{-1}=P$. This Hamiltonian
operator satisfies the conditions discussed in the preceding section and is
PT-symmetric when $\lambda =ig$, $g$ real, because $V(-x)^{*}=V(x)$ (see,
for example, \cite{B07} and references therein).

Ahmed\cite{A09} proved that when $\lambda =ig$ the eigenvalues are solutions
to the nonlinear equation
\begin{equation}
D(E,g)=\Re \left[ \frac{\left( \frac{1}{4}-ig\right) ^{1/4}}{\Gamma \left(
\frac{3}{4}-\frac{E}{2\sqrt{1+4ig}}\right) \Gamma \left( \frac{1}{4}-\frac{E%
}{2\sqrt{1-4ig}}\right) }\right] =0,  \label{eq:D(E,g)}
\end{equation}
and from a straightforward numerical calculation he conjectured that the
number of real eigenvalues decreases from infinity ($g=0$) to just one (for
a sufficiently large value of $g$).

The reason for this behaviour is that as $g$ increases from $g=0$ a pair of
eigenvalues $E_{2n-1}$ and $E_{2n}$, $n=1,2,\ldots $, approach each other,
coalesce at $g=g_{n}>0$ and become a pair of complex conjugate numbers for $%
g>g_{n}$. The only eigenvalue that appears to remain real for all $g$ is $%
E_{0}$. This result is consistent with the fact that there is just one real
eigenvalue when $V(x)=i|x|x$\cite{FS14} that is the potential for the
strong-coupling limit ($g\rightarrow \infty $) of (\ref{eq:H_model1}). In
fact, from the canonical transformation
\begin{equation}
U_{\gamma }xU_{\gamma }^{-1}=\gamma x,U_{\gamma }pU_{\gamma }^{-1}=\gamma
^{-1}p  \label{eq:Can_transf}
\end{equation}
with $\gamma =g^{-1/4}$ we easily prove that
\begin{equation}
\lim\limits_{g\rightarrow \infty }g^{-1/2}U_{\gamma }xU_{\gamma
}^{-1}=p^{2}+i|x|x.  \label{eq:strong_coupling}
\end{equation}

\begin{figure}[H]
\begin{center}
\includegraphics[width=9cm]{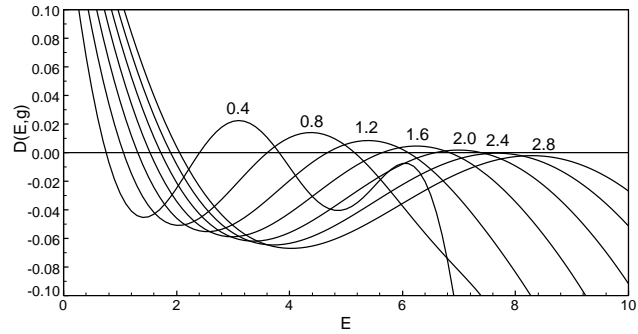}
\end{center}
\caption{$D(E,g)$ vs. $E$ for $g=0.4,\ 0.8,\ 1.2,\ 1.6,\ 2.0,\ 2.4,\ 2.8$}
\label{fig:Coalescence}
\end{figure}

Numerical calculation shows that the critical points $g_{n}$ decrease with $%
n $ so that for a given value of $g$ only the pairs of eigenvalues with $%
g_{n}<g$ are real. When $g>g_{1}$ there is only one real eigenvalue ($E_{0}$%
) and when $g_{n+1}<g<g_{n}$ there are $2n+1$ real eigenvalues. This fact
explains why Ahmed\cite{A09} obtained an odd number of real eigenvalues for
every $g<g_{1}$. For example, Fig.~\ref{fig:Coalescence} illustrates the
coalescence of the eigenvalues $E_{1}$ and $E_{2}$. Ahmed\cite{A09} suggests
that there is just one real eigenvalue for $g=2$; however, our calculation
shows that there are three real eigenvalues: $E_{0}=1.720857958$, $%
E_{1}=6.579362071$ and $E_{2}=7.39812626$. He probably missed the two
eigenvalues $E_{1}$ and $E_{2}$ because they are quite close to each other
or due to insufficient accuracy in the calculation.

Equation (\ref{eq:D(E,g)}) yields either $E(g)$ or $g(E)$. If we take into
account that $dD/dE=0$ when $E$ and $g$ are linked by $D(E,g)=0$ then we
find that
\begin{equation}
\frac{dg}{dE}=\frac{\partial D/\partial E}{\partial D/\partial g}.
\end{equation}
Therefore, the coalescence points that are given by $dg/dE=0$ (square-root
branch points) can be easily obtained by solving the system of nonlinear
equations (see, for example, \cite{HS90,FG14c})
\begin{equation}
D(E,g)=0,\;\frac{\partial D}{\partial E}(E,g)=0.  \label{eq:crit-point-eqs}
\end{equation}
Table~\ref{tab:CP1} shows the first five critical points (note that $g_{1}>2$%
).

As argued in the preceding section, the application of perturbation theory
to the model (\ref{eq:H_model1}) should yield a $g^{2}$-series for every
eigenvalue $E_{n}(ig)$. Since both $H(0)$ and $dH(\lambda )/d\lambda =|x|x$
are real, then the coefficients $E_{n}^{(j)}$ are real. The first three
terms of the perturbation series for the first three eigenvalues are:

\begin{eqnarray}
E_{0}(ig) &=&\frac{1}{2}+\left( 2+\ln 2\right) {g}^{2}+  \nonumber \\
&&\left[ -7\,\ln 2+3\,\left( \ln 2\right) ^{2}+\frac{1}{3}\,\left( \ln
2\right) ^{3}-\frac{1}{4}\,\zeta \left( 3\right) -12+\frac{1}{12}\,{\pi }%
^{2}\ln 2-\frac{1}{12}\,{\pi }^{2}\right] {g}^{4}+\ldots  \nonumber \\
&=&0.5000000000+2.693147181g^{2}-15.85255355g^{4}+\ldots
\label{eq:E0_series}
\end{eqnarray}

\begin{eqnarray}
E_{1}(ig) &=&\frac{3}{2}+\left( 3+9\,\ln 2\right) {g}^{2}+  \nonumber \\
&&\left[ -{\frac{81}{4}}\,\zeta \left( 3\right) -63\,\ln 2+54\,\left( \ln
2\right) ^{2}-18+{\frac{27}{4}}\,{\pi }^{2}\ln 2+27\,\left( \ln 2\right)
^{3}-\frac{9}{2}\,{\pi }^{2}\right] {g}^{4}+\ldots  \nonumber \\
&=&1.500000000+9.238324625g^{2}-49.30966898g^{4}+\ldots  \label{eq:E1_series}
\end{eqnarray}

\begin{eqnarray}
E_{2}(ig) &=&\frac{5}{2}+\left( -\frac{5}{2}+25\,\ln 2\right) {g}^{2}+
\nonumber \\
&&\left[ -{\frac{625}{4}}\,\zeta \left( 3\right) +75\,\ln 2+{\frac{125}{2}}%
\,\left( \ln 2\right) ^{2}+{\frac{625}{12}}\,{\pi }^{2}\ln 2-{\frac{405}{8}}+%
{\frac{625}{3}}\,\left( \ln 2\right) ^{3}-{\frac{875}{24}}\,{\pi }%
^{2}\right] {g}^{4}+\ldots  \nonumber \\
&=&2.500000000+14.82867952g^{2}-90.5745397g^{2}+\ldots  \label{eq:E2_series}
\end{eqnarray}
where $\zeta \left( k\right) $ is the zeta function. The radius of
convergence of the perturbation series for the pair of eigenvalues $E_{2n-1}$
and $E_{2n}$, $n>0$ is expected to be $g_{n}$.

The eigenvalues for the second model
\begin{equation}
H(\lambda )=p^{2}+|x|+\lambda x,  \label{eq:H_model2}
\end{equation}
when $\lambda =ig$, are roots of the nonlinear expression\cite{A09}
\begin{eqnarray}
D(E,g) &=&\nu _{1}\mu _{2}Ai(E/\mu _{2})Ai^{\prime }(E/\mu _{1})-\mu _{1}\nu
_{2}Ai(E/\mu _{1})Ai^{\prime }(E/\mu _{2})=0,  \nonumber \\
\nu _{1} &=&-1+ig,\;\nu _{2}=1+ig,\;\mu _{1}=-\left( \nu _{1}^{2}\right)
^{1/3},\;\mu _{2}=-\left( \nu _{2}^{2}\right) ^{1/3}.  \label{eq:D(E,g)_2}
\end{eqnarray}
The behaviour of the spectrum of the PT-symmetric Hamiltonian (\ref
{eq:H_model2}) is similar to the one discussed above in all relevant aspects
except one. In this case the strong-coupling limit is given by the canonical
transformation (\ref{eq:Can_transf}) with $\gamma =g^{-1/3}$ and
\begin{equation}
\lim\limits_{g\rightarrow \infty }g^{-2/3}U_{\gamma }HU_{\gamma
}^{-1}=p^{2}+ix.  \label{eq:strong_coup_2}
\end{equation}
Bender and Boettcher\cite{BB98} proved that this model does not exhibit
eigenvalues and that the ground state of $H=p^{2}-(ix)^{N}$ diverges as $%
N\rightarrow 1^{+}$. This is the most relevant difference with respect to
the preceding example.

The first four critical points of the spectrum of (\ref{eq:H_model2}) are
shown in Table~\ref{tab:CP2}.

\section{Conclusions}

\label{sec:conclusions}

The two PT-symmetric Hamiltonians discussed in the preceding section exhibit
critical points $g_{n}>0$ that decrease with $n$ (we omit the consideration
of the case $g<0$ because $E(-ig)=E(ig)$). Since it appears that $%
\lim\limits_{n\rightarrow \infty }g_{n}=0$ the number of real eigenvalues is
finite for all $g>0$. This result is important because it may shed light on
the spectra of a family of PT-symmetric multidimensional oscillators with
critical points that decrease with the magnitude of the coalescing
eigenvalues\cite{BW12,AFG14c}. Present results suggest that the PT-phase
transition\cite{BW12} for the Hamiltonians (\ref{eq:H_model1}) and (\ref
{eq:H_model2}) takes place at the trivial Hermitian limit $g=0$. Such a
behaviour commonly takes place in multidimensional problems that exhibit
some kind of point-group symmetry\cite{AFG14c,FG14a,FG14b,FG15}. For this
reason the results above for one-dimensional models that merely exhibit
parity symmetry $PH(0)P=H(0)$ appear to be quite interesting.

\begin{table}[H]
\caption{First critical points for the Hamiltonian (\ref{eq:H_model1})}
\label{tab:CP1}
\begin{center}
\begin{tabular}{D{.}{.}{2}D{.}{.}{20}D{.}{.}{20}}
\hline
 \multicolumn{1}{c}{$n$} & \multicolumn{1}{c}{$g_n$} &\multicolumn{1}{c}{$E_{2n-1}=E_{2n}$}\\
\hline
 1 &  2.3262718829432516237   &     7.5466636499753202587 \\
 2 &  0.38031593803698939339  &     5.9543772352940988301 \\
 3 &  0.24202011969988939326  &     7.5185674146488458902 \\
 4 &  0.18511586549014097563  &     9.3051810150783521871 \\
 5 &  0.15263064129841349291  &    11.167144554229643582  \\
 \end{tabular}
\end{center}
\end{table}

\begin{table}[H]
\caption{First critical points for the Hamiltonian (\ref{eq:H_model2})}
\label{tab:CP2}
\begin{center}
\begin{tabular}{D{.}{.}{2}D{.}{.}{20}D{.}{.}{20}}
\hline
 \multicolumn{1}{c}{$n$} & \multicolumn{1}{c}{$g_n$} &\multicolumn{1}{c}{$E_{2n-1}=E_{2n}$}\\
\hline
 1 &  0.62782075783846150477  &     3.3482781381164933022 \\
 2 &  0.35319223654461721210  &     4.7569117414811876143 \\
 3 &  0.25419305000674157648  &     6.0526901706522118370 \\
 4 &  0.20120031566314106331  &     7.2414114087019937638 \\
 \end{tabular}
\end{center}
\end{table}

\end{document}